\documentclass[10pt]{article}
\usepackage{graphicx}
\usepackage{amsmath}
\usepackage{amssymb}
\usepackage{caption2}
\begin{document}
\begin{center}
{\large {\bf \sc{   Triply-charmed hexaquark states with the QCD sum rules
  }}} \\[2mm]
Zhi-Gang  Wang \footnote{E-mail: zgwang@aliyun.com.  }   \\
 Department of Physics, North China Electric Power University, Baoding 071003, P. R. China
\end{center}

\begin{abstract}
In this article,  we construct the charmed-diquark-charmed-diquark-charmed-diquark type current to study the axialvector triply-charmed  hexaquark state  with the QCD sum rules in details. In calculations, we take the energy scale formula  $\mu=\sqrt{M^2_{H}-(3{\mathbb{M}}_c)^2}$ to choose the pertinent  energy scale of the QCD spectral density so as to  enhance the pole contribution and improve the convergent behavior of the operator product expansion.
  If the spin-breaking effects are small for the triply-charmed  hexaquark states,  the ground state hexaquark states with $J^P=0^+$, $1^+$ and $2^+$ are estimated to have the masses about $5.8\,\rm{GeV}$ and narrow widths.
\end{abstract}

PACS number: 12.39.Mk, 12.38.Lg

Key words: Hexaquark states, QCD sum rules

\section{Introduction}

A number of charmonium-like and bottomonium-like states were observed after the observation  of the $X(3872)$, the most elusive meson up to now,  by the  Belle collaboration \cite{X3872-2003}. It is very difficult to find rooms to accommodate those exotic $X$, $Y$ and $Z$ states in the $q\bar{q}$ meson spectrum comfortably even for the charge-neutral mesons, such as the $Y(4260)$, $Y(4360)$, $Y(4660)$, etc.  The charged
charmonium-like states and bottomonium-like states are very  good candidates for the hidden-charm and hidden-bottom   tetraquark states or molecular states \cite{HXChen-review-1601, RFLebed-review-1610,AEsposito-review-1611,FKGuo-review-1705, AAli-review-1706,SLOlsen-review-1708,MNielsen-review-1812,YRLiu-review-1903,CPShen-review-1907}.
The QCD sum rules play  an important role  in diagnosing the nature of   those new charmonium-like states \cite{MNielsen-review-1812,QCDSR-4-quark-mass,WangHuangtao-PRD,Wang-tetra-formula,ZGWang-CTP,QCDSR-4-quark-width,WangZG-4-quark-mole}.

Now let us discuss how to construct the interpolating currents  to study the tetraquark states in the QCD sum rules.
The scattering amplitude for one-gluon exchange  is proportional to
\begin{eqnarray}
\left(\frac{\lambda^a}{2}\right)_{ij}\left(\frac{\lambda^a}{2}\right)_{kl}&=&-\frac{N_c+1}{4N_c}t^A_{ik}t^A_{lj}+\frac{N_c-1}{4N_c}t^S_{ik}t^S_{lj} \, ,
\end{eqnarray}
where
\begin{eqnarray}
t^A_{ik}t^A_{lj}&=&\delta_{ij}\delta_{kl}-\delta_{il}\delta_{kj}=\varepsilon_{mik}\varepsilon_{mjl}\, , \nonumber\\
t^S_{ik}t^S_{lj}&=&\delta_{ij}\delta_{kl}+\delta_{il}\delta_{kj}\, ,
\end{eqnarray}
the $\lambda^a$ is the  Gell-Mann matrix,  the $i$, $j$, $k$, $m$ and $l$ are color indexes, the $N_c$ is the color number.  The negative sign in front of the $t^A_{ik}t^A_{lj}$ represents the interaction
is attractive and favors  forming  the diquark correlations in  color antitriplet,   the positive sign in front of the  $t^S_{ik}t^S_{lj}$ represents
 the interaction  is repulsive and  disfavors  forming  the diquark correlations in  color sextet.

The   diquark operators $\varepsilon^{ijk} q^{T}_j C\Gamma q^{\prime}_k$ in color antitriplet
  have  five  structures, where $C\Gamma=C\gamma_5$, $C$, $C\gamma_\mu \gamma_5$,  $C\gamma_\mu $ and $C\sigma_{\mu\nu}$ correspond to the scalar, pseudoscalar, vector, axialvector  and  tensor diquarks, respectively.
The  QCD sum rules  indicate that  the favored quark-quark configurations are the scalar and axialvector diquark states,  the axialvector diquark states have slightly larger masses than the corresponding scalar diquark states \cite{WangDiquark,WangLDiquark}, in the case of the heavy-light diquark states, they have almost  degenerated  masses  \cite{WangDiquark}. In the QCD sum rules, we usually choose the scalar and axialvector diquark operators to construct the tetraquark current operators to interpolate the diquark-antidiquark type tetraquark states with  the lowest masses. For example, we study  the $Z_c(3900)$
with the  $C\gamma_5\otimes \gamma_\mu C-C\gamma_\mu\otimes\gamma_5 C$ type tetraquark current \cite{WangHuangtao-PRD}.
The masses and decay widths of the diquark-antidiquark type tetraquark states have been studied extensively with the QCD sum rules \cite{MNielsen-review-1812,QCDSR-4-quark-mass,WangHuangtao-PRD,Wang-tetra-formula,ZGWang-CTP,QCDSR-4-quark-width}.

In previous works, we  studied the energy scale dependence of the QCD sum rules for the exotic $X$, $Y$, $Z$ states, which are very good candidates for the  hidden-charm and hidden-bottom tetraquark states and molecular states,  for the first time, and suggested a  formula,
\begin{eqnarray}
\mu&=&\sqrt{M^2_{X/Y/Z}-(2{\mathbb{M}}_Q)^2} \, ,
 \end{eqnarray}
 with the effective heavy quark mass ${\mathbb{M}}_Q$  to choose the best  energy scales of the  QCD spectral densities \cite{WangHuangtao-PRD,Wang-tetra-formula,ZGWang-CTP,WangZG-4-quark-mole}. In calculations, we observe that the energy scalar formula can enhance the pole contributions remarkably and improve the convergent behaviors of the operator product expansion remarkably also.

In 2017, the LHCb collaboration observed the doubly charmed baryon state  $\Xi_{cc}^{++}$ in the $\Lambda_c^+ K^- \pi^+\pi^+$ mass spectrum  \cite{LHCb-Xicc}.
The observation of the $\Xi_{cc}^{++}$  provides  valuable  experimental information on the strong correlation between the two charm quarks, which maybe
 shed light on the
  spectroscopy of the doubly-charmed  baryon states, tetraquark states,  pentaquark states and hexaquark states.
     For the heavy-quark-heavy-quark systems  $QQ$,  only the   axialvector  diquark operators $\varepsilon^{ijk} Q^{T}_j C\gamma_\mu Q_k$ and tensor diquark operators $\varepsilon^{ijk} Q^{T}_j C\sigma_{\mu\nu }Q_k$ can exist  due to the Fermi-Dirac  statistics,
we usually take the axialvector diquark operators $\varepsilon^{ijk}  Q^T_jC\gamma_\mu Q_k$ as the basic constituents to construct the four-quark currents to study the doubly heavy  tetraquark states with the QCD sum rules \cite{Nielsen-Lee,QQ-QCDSR-Wang2011,QQ-QCDSR-Chen,Wang-QQ-tetraquark}. For the doubly heavy hexaquark states, we can choose the doubly heavy diquark operators  $\varepsilon^{ijk}  Q^T_jC\gamma_\mu Q_k$ or heavy diquark operators $\varepsilon^{ijk}  q^T_jC\gamma_5 Q_k$ and $\varepsilon^{ijk}  q^T_jC\gamma_\mu Q_k$ as the basic constituents to construct the six-quark currents.
In Ref.\cite{WZG-hexaQ}, we extend our previous works  to study the scalar-diquark-scalar-diquark-scalar-diquark ($\varepsilon^{aij}u^{T}_iC\gamma_5d_j-\varepsilon^{bkl}u^{T}_kC\gamma_5 c_l-\varepsilon^{cmn} d^{T}_mC\gamma_5 c_n$) type hexaquark state $uuddcc$ with the QCD sum rules in details.  In Ref.\cite{WZG-ccc-dibaryon}, we construct the color-singlet-color-singlet type currents to study the scalar and axialvector (triply-charmed) $\Xi_{cc}\Sigma_c$ dibaryon states with QCD sum rules in details.
In this article, we extend our previous works  to study the charmed-diquark-charmed-diquark-charmed-diquark type hexaquark states with the QCD sum rules.

The article is arranged as follows:  we derive the QCD sum rules for the mass and pole residue of  the
axialvector  triply-charmed hexaquark state in Sect.2;  in Sect.3, we present the numerical results and discussions; and Sect.4 is reserved for our
conclusion.

\section{QCD sum rules for  the triply-charmed hexaquark states}

We choose the scalar ($S$) and axialvector ($A$) charmed diquark operators,
\begin{eqnarray}
S^i_{qc}&=&\varepsilon^{ijk}q^T_j(x) C\gamma_5 c_k(x)\, ,\nonumber\\
A^i_{qc,\mu}&=&\varepsilon^{ijk}q^T_j(x) C\gamma_\mu c_k(x)\, ,
\end{eqnarray}
as the basic constituents to construct the triply-charmed hexaquark currents.
 For the spin $J=0$ hexaquark states, we can construct  two currents $J(x)$,
 \begin{eqnarray}
J(x)&=&\varepsilon^{ijk} S^i_{uc}\,S^j_{dc}\,S^k_{sc}\,\, , \, \, \varepsilon^{ijk} A^i_{qc,\mu}\,A^j_{q^{\prime}c,\nu}\,S^k_{\tilde{q}c}\,g^{\mu\nu}\, ,
\end{eqnarray}
with $q,\,q^{\prime},\,\tilde{q}=u,\,d,\,s$ and $q\neq q^{\prime}$.
For  the spin $J=1$ hexaquark states, we can construct  three currents $J_{\mu/\mu\nu}(x)$,
 \begin{eqnarray}
J_{\mu/\mu\nu}(x)&=&\varepsilon^{ijk} S^i_{qc}\,S^j_{q^{\prime}c}\,A^k_{\tilde{q}c,\mu}\,\, , \, \, \varepsilon^{ijk} A^i_{qc,\alpha}\,A^j_{\hat{q}c,\beta}\,S^k_{\tilde{q}c}\,\left(g^{\mu\alpha}g^{\nu\beta}-g^{\mu\beta}g^{\nu\alpha}\right)\, ,\,\nonumber\\
&&\varepsilon^{ijk} A^i_{qc,\alpha}\,A^j_{\hat{q}c,\beta}\,A^k_{\tilde{q}c,\sigma}\,\varepsilon^{\alpha\beta\sigma\mu}\, ,
\end{eqnarray}
with $q,\,q^{\prime},\,\tilde{q},\,\hat{q}=u,\,d,\,s$ and $q\neq q^{\prime}$.
For  the spin $J=2$ hexaquark states, we can construct  two currents $J_{\mu\nu/\mu\nu\alpha}(x)$,
 \begin{eqnarray}
J_{\mu\nu/\mu\nu\alpha}(x)&=& \varepsilon^{ijk} A^i_{qc,\alpha}\,A^j_{\hat{q}c,\beta}\,S^k_{\tilde{q}c}\,\left(g^{\mu\alpha}g^{\nu\beta}+g^{\mu\beta}g^{\nu\alpha}\right)\, ,\,\nonumber\\
&&\varepsilon^{ijk} A^i_{qc,\mu}\,A^j_{\hat{q}c,\nu}\,A^k_{\tilde{q}c,\alpha}+\left(\mu\nu\alpha\to \nu\mu\alpha\, ,\, -\alpha\mu\nu\, ,\, -\alpha\nu\mu\right)\, ,
\end{eqnarray}
with $q,\,\hat{q},\,\tilde{q}=u,\,d,\,s$, $q\neq \hat{q}$.
For  the spin $J=3$ hexaquark states, we can construct  one current $J_{\mu\nu\alpha}(x)$,
 \begin{eqnarray}
J_{\mu\nu\alpha}(x)&=& \varepsilon^{ijk} A^i_{uc,\mu}\,A^j_{dc,\nu}\,A^k_{sc,\alpha}+\left(\mu\nu\alpha\to \mu\alpha\nu\, ,\, \alpha\mu\nu\, ,\, \alpha\nu\mu\, , \, \nu\mu\alpha\, , \nu\alpha\mu\right)\, .
\end{eqnarray}
In this article, we take the isospin limit, and intend to estimate the masses of the  lowest triply charmed hexaquark states.   As the $s$-quark has larger mass due to the flavor $SU(3)$ breaking effect, we retain  the three currents
 $J(x)$, $J_{\mu}(x)$ and $J_{\mu\nu}(x)$ without the $s$-quark operators,
 \begin{eqnarray}
 J(x)&=&  \varepsilon^{abc}\varepsilon^{aij}\varepsilon^{bkl}\varepsilon^{cmn}u^T_i(x)C\gamma_{\alpha}c_j(x) d^T_k(x)C\gamma^{\alpha}c_l(x)u^T_m(x)C\gamma_{5}c_n(x)\, , \nonumber\\
 J_\mu(x)&=&  \varepsilon^{abc}\varepsilon^{aij}\varepsilon^{bkl}\varepsilon^{cmn}u^T_i(x)C\gamma_5c_j(x) d^T_k(x)C\gamma_5c_l(x)u^T_m(x)C\gamma_{\mu}c_n(x)\, , \nonumber\\
  J_{\mu\nu}(x)&=&  \varepsilon^{abc}\varepsilon^{aij}\varepsilon^{bkl}\varepsilon^{cmn} u^T_i(x)C\gamma_{\mu}c_j(x) d^T_k(x)C\gamma_{\nu}c_l(x)u^T_m(x)C\gamma_{5}c_n(x)+\left(\mu\leftrightarrow\nu\right)\, ,
\end{eqnarray}
where the $a$, $b$, $c$, $\cdots$ are color indexes.

Those diquark-diquark-diquark type currents couple potentially to the hexaquark states ($H$) with the spin-parity $J^P=0^+$, $1^+$ and $2^+$, respectively,
\begin{eqnarray}
\langle 0|J(0)|H_0(p)\rangle &=&\lambda_{H_0}\, ,\nonumber\\
\langle 0|J_\mu(0)|H_1(p)\rangle &=&\lambda_{H_1}\,\varepsilon_\mu\, ,\nonumber\\
\langle 0|J_{\mu\nu}(0)|H_2(p)\rangle &=&\lambda_{H_2}\,\varepsilon_{\mu\nu}\, ,
\end{eqnarray}
where the $\lambda_{H_{0/1/2}}$ are the pole residues,  the $\varepsilon_\mu$ and $\varepsilon_{\mu\nu}$ are the polarization vectors of the axialvector and tensor hexaquark states, respectively.
The $H$'s have three diquarks, $H_0=A_{uc}\,A_{dc}\,S_{uc}$, $H_1=S_{uc}\,S_{dc}\,A_{uc}$, $H_2=A_{uc}\,A_{dc}\,S_{uc}$.
The axialvector charmed diquark states have slightly larger masses than the scalar charmed diquark states, or they have almost  degenerated  masses  \cite{WangDiquark},
the masses of the triply-charmed hexaquark states maybe have the hierarchy  $M_{1}\leq M_{0}\leq M_{2}$. It is horrible to carry out the operator product expansion for the triply heavy hexaquark states, in this article, we choose the axialvector current $J_\mu(x)$ to study the lowest state $H_1$ to estimate the magnitude   of the masses of the triply-charmed hexaquark states, and the conclusion should be taken with caution, as  we could obtain more robust  predictions  by choosing the most general currents.

In the following, we write down  the two-point correlation function  $\Pi_{\mu\nu}(p)$ in the QCD sum rules,
\begin{eqnarray}
\Pi_{\mu\nu}(p)&=&i\int d^4x e^{ip \cdot x} \langle0|T\Big\{J_{\mu}(x)J^{\dagger}_{\nu}(0)\Big\}|0\rangle \, .
\end{eqnarray}

At the hadron side of the correlation function  $\Pi_{\mu\nu}(p)$,   we  isolate  the contribution of   the lowest axialvector triply-charmed hexaquark state,
\begin{eqnarray}
\Pi_{\mu\nu}(p)&=& \frac{\lambda_{H}^2}{M_{H}^2-p^2}\left( -g_{\mu\nu}+\frac{p_{\mu}p_{\nu}}{p^2}\right)+\cdots\, ,   \nonumber\\
&=&\Pi(p^2)\left( -g_{\mu\nu}+\frac{p_{\mu}p_{\nu}}{p^2}\right)+\cdots\, ,
\end{eqnarray}
thereafter we will smear the subscript $1$.
In this article, we choose the tensor structure $-g_{\mu\nu}+\frac{p_{\mu}p_{\nu}}{p^2}$  and study the component $\Pi(p^2)$ to explore the axialvector hexaquark state.

At the QCD side of the correlation function  $\Pi_{\mu\nu}(p)$, we contract the $u$, $d$ and $c$ quark fields  with Wick theorem and obtain the result,
\begin{eqnarray}\label{PI-A}
\Pi_{\mu\nu}(p)&=&i\,\varepsilon^{abc}\varepsilon^{aij}\varepsilon^{bkl}\varepsilon^{cmn}  \varepsilon^{a^{\prime}b^{\prime}c^{\prime}}\varepsilon^{a^{\prime}i^{\prime}j^{\prime}}\varepsilon^{b^{\prime}k^{\prime}l^{\prime}}\varepsilon^{c^{\prime}m^{\prime}n^{\prime}} \int d^4x\, e^{ip\cdot x} \nonumber\\
&&\Big\{  {\rm Tr}\left[\gamma_5  C_{jj^\prime}(x)   \gamma_{5} C U_{ii^\prime }^T(x)C\right] \,{\rm Tr}\left[\gamma_5  C_{ll^\prime}(x)   \gamma_{5} C D^T_{kk^\prime }(x)C\right]   {\rm Tr} \left[\gamma_\mu C_{nn^\prime}(x) \gamma_\nu C U^T_{mm^\prime }(x)C \right]       \nonumber\\
&&+  {\rm Tr}\left[\gamma_5  C_{jn^\prime}(x)   \gamma_{\nu} C U_{im^\prime }^T(x)C\right] \,{\rm Tr}\left[\gamma_5  C_{ll^\prime}(x)   \gamma_{5} C D^T_{kk^\prime }(x)C\right]   {\rm Tr} \left[\gamma_\mu C_{nj^\prime}(x) \gamma_5 C U^T_{mi^\prime }(x)C \right]      \nonumber\\
&&- {\rm Tr}\left[\gamma_5  C_{ll^\prime}(x)   \gamma_{5} C D_{kk^\prime }^T(x)C\right] \,{\rm Tr}\left[\gamma_5  C_{jj^\prime}(x)   \gamma_{5} C U^T_{mi^\prime }(x)C \gamma_\mu C_{nn^\prime}(x) \gamma_\nu C U^T_{im^\prime }(x)C \right]      \nonumber\\
&&- {\rm Tr}\left[\gamma_5  C_{jj^\prime}(x)   \gamma_{5} C U_{ii^\prime }^T(x)C\right] \,{\rm Tr}\left[\gamma_5  C_{ln^\prime}(x)   \gamma_{\nu} C U^T_{mm^\prime }(x)C \gamma_\mu C_{nl^\prime}(x) \gamma_5 C D^T_{kk^\prime }(x)C \right]      \nonumber\\
&&- {\rm Tr}\left[\gamma_5  C_{ll^\prime}(x)   \gamma_{5} C D_{kk^\prime }^T(x)C\right] \,{\rm Tr}\left[\gamma_5  C_{jn^\prime}(x)   \gamma_{\nu} C U^T_{mm^\prime }(x)C \gamma_\mu C_{nj^\prime}(x) \gamma_5 C U^T_{ii^\prime }(x)C \right]      \nonumber\\
&&- {\rm Tr}\left[\gamma_\mu  C_{nn^\prime}(x)   \gamma_{\nu} C U_{mm^\prime }^T(x)C\right] \,{\rm Tr}\left[\gamma_5  C_{jl^\prime}(x)   \gamma_{5} C D^T_{kk^\prime }(x)C \gamma_5 C_{lj^\prime}(x) \gamma_5 C U^T_{ii^\prime }(x)C \right]      \nonumber\\
&&- {\rm Tr}\left[\gamma_\mu  C_{nj^\prime}(x)   \gamma_{5} C U_{mi^\prime }^T(x)C\right] \,{\rm Tr}\left[\gamma_5  C_{jl^\prime}(x)   \gamma_{5} C D^T_{kk^\prime }(x)C \gamma_5 C_{ln^\prime}(x) \gamma_\nu C U^T_{im^\prime }(x)C \right]      \nonumber\\
&&- {\rm Tr}\left[\gamma_5  C_{jn^\prime}(x)   \gamma_{\nu} C U_{im^\prime }^T(x)C\right] \,{\rm Tr}\left[\gamma_5  C_{lj^\prime}(x)   \gamma_{5} C U^T_{mi^\prime }(x)C \gamma_\mu C_{nl^\prime}(x) \gamma_5 C D^T_{kk^\prime }(x)C \right]      \nonumber\\
&&+ {\rm Tr}\left[\gamma_5  C_{jj^\prime}(x)   \gamma_{5} C U_{mi^\prime }^T(x)C\gamma_\mu  C_{nl^\prime}(x)   \gamma_{5} C D^T_{kk^\prime }(x)C \gamma_5
 C_{ln^\prime}(x) \gamma_\nu C U^T_{im^\prime }(x)C \right]      \nonumber\\
 &&+ {\rm Tr}\left[\gamma_5  C_{jl^\prime}(x)   \gamma_{5} C D_{kk^\prime }^T(x)C\gamma_5  C_{lj^\prime}(x)   \gamma_{5} C U^T_{mi^\prime }(x)C \gamma_\mu
 C_{nn^\prime}(x) \gamma_\nu C U^T_{im^\prime }(x)C \right]      \nonumber\\
 &&+ {\rm Tr}\left[\gamma_5  C_{jl^\prime}(x)   \gamma_{5} C D_{kk^\prime }^T(x)C\gamma_5  C_{ln^\prime}(x)   \gamma_{\nu} C U^T_{mm^\prime }(x)C \gamma_\mu
 C_{nj^\prime}(x) \gamma_5 C U^T_{ii^\prime }(x)C \right]      \nonumber\\
&& + {\rm Tr}\left[\gamma_5  C_{jn^\prime}(x)   \gamma_{\nu} C U_{mm^\prime }^T(x)C\gamma_\mu  C_{nl^\prime}(x)   \gamma_{5} C D^T_{kk^\prime }(x)C \gamma_5
 C_{lj^\prime}(x) \gamma_5 C U^T_{ii^\prime }(x)C \right]  \Big\}\, ,
\end{eqnarray}
where $S_{ij}(x)=U_{ij}(x)$ and $D_{ij}(x)$,
\begin{eqnarray}\label{LQuarkProg}
S_{ij}(x)&=& \frac{i\delta_{ij}\!\not\!{x}}{ 2\pi^2x^4}-\frac{\delta_{ij}\langle
\bar{q}q\rangle}{12} -\frac{\delta_{ij}x^2\langle \bar{q}g_s\sigma Gq\rangle}{192} -\frac{ig_sG^{a}_{\alpha\beta}t^a_{ij}(\!\not\!{x}
\sigma^{\alpha\beta}+\sigma^{\alpha\beta} \!\not\!{x})}{32\pi^2x^2}  \nonumber\\
&& -\frac{\delta_{ij}x^4\langle \bar{q}q \rangle\langle g_s^2 GG\rangle}{27648} -\frac{1}{8}\langle\bar{q}_j\sigma^{\mu\nu}q_i \rangle \sigma_{\mu\nu}+\cdots \, ,
\end{eqnarray}
\begin{eqnarray}\label{HQuarkProg}
C_{ij}(x)&=&\frac{i}{(2\pi)^4}\int d^4k e^{-ik \cdot x} \left\{
\frac{\delta_{ij}}{\!\not\!{k}-m_c}
-\frac{g_sG^n_{\alpha\beta}t^n_{ij}}{4}\frac{\sigma^{\alpha\beta}(\!\not\!{k}+m_c)+(\!\not\!{k}+m_c)
\sigma^{\alpha\beta}}{(k^2-m_c^2)^2}\right.\nonumber\\
&&\left. -\frac{g_s^2 (t^at^b)_{ij} G^a_{\alpha\beta}G^b_{\mu\nu}(f^{\alpha\beta\mu\nu}+f^{\alpha\mu\beta\nu}+f^{\alpha\mu\nu\beta}) }{4(k^2-m_c^2)^5}+\cdots\right\} \, ,\nonumber\\
f^{\alpha\beta\mu\nu}&=&(\!\not\!{k}+m_c)\gamma^\alpha(\!\not\!{k}+m_c)\gamma^\beta(\!\not\!{k}+m_c)\gamma^\mu(\!\not\!{k}+m_c)\gamma^\nu(\!\not\!{k}+m_c)\, ,
\end{eqnarray}
and  $t^n=\frac{\lambda^n}{2}$, the $\lambda^n$ is the Gell-Mann matrix
\cite{WangHuangtao-PRD,Reinders85,Pascual-1984}.
In the full light quark propagator, see Eq.\eqref{LQuarkProg}, we add the term $\langle\bar{q}_j\sigma_{\mu\nu}q_i \rangle$,  which comes  from  Fierz rearrangement  of the quark-antiquark pair $\langle q_i \bar{q}_j\rangle$ to  absorb the gluons  emitted from other  quark lines,  to extract  the mixed condensates   $\langle\bar{q}g_s\sigma G q\rangle$, $\langle\bar{q}g_s\sigma G q\rangle^2$ and $\langle\bar{q}g_s\sigma G q\rangle^3$, respectively \cite{WangHuangtao-PRD}.
There are three light quark lines (or propagators) and three heavy quark lines (or propagators) in the correlation function $\Pi_{\mu\nu}(p)$, see Eq.\eqref{PI-A}, if each heavy quark line
emits a gluon and each light quark line contributes quark-antiquark pair, we obtain a quark-gluon  operator $g_sG_{\mu\nu}g_sG_{\alpha\beta}g_sG_{\lambda\tau}\bar{q}q\bar{q}q\bar{q}q$, which is of dimension 15, and leads to the vacuum condensates
$\langle\frac{\alpha_sGG}{\pi}\rangle\langle\bar{q}q\rangle^2\langle\bar{q}g_s\sigma Gq\rangle$, $\langle g_s^3 GGG\rangle\langle\bar{q}q\rangle^3$ and $\langle\bar{q}g_s\sigma Gq\rangle^3$.

In the QCD sum rules for the tetraquark (molecular) states, pentaquark (molecular) states and hexaquark states (or dibaryon states), we take into account the vacuum condensates,  which are vacuum expectations of the quark-gluon operators of the order $\mathcal{O}(\alpha_s^k)$ with $k\leq1$ in a consistent way \cite{WangHuangtao-PRD,Wang-tetra-formula,ZGWang-CTP,Wang-QQ-tetraquark,WZG-hexaQ,WZG-ccc-dibaryon,QCDSR-WangZG-5-quark-mole,QCDSR-5-quark-penta}. In the present case, if we take the truncation $k\leq1$, the highest dimensional vacuum condensates are $\langle\bar{q}q\rangle^3\langle\frac{\alpha_sGG}{\pi}\rangle$ and $\langle\bar{q} q\rangle\langle\bar{q}g_s\sigma Gq\rangle^2$,  the vacuum condensates $\langle\frac{\alpha_sGG}{\pi}\rangle\langle\bar{q}q\rangle^2\langle\bar{q}g_s\sigma Gq\rangle$, $\langle g_s^3 GGG\rangle\langle\bar{q}q\rangle^3$ and $\langle\bar{q}g_s\sigma Gq\rangle^3$  come from the quark-gluon operators of the order $\mathcal{O}(\alpha_s^{\frac{3}{2}})$ and should be discarded. In this article, we take into account the vacuum condensate $\langle\bar{q}g_s\sigma Gq\rangle^3$ and neglect the vacuum condensates $\langle g_s^3 GGG\rangle\langle\bar{q}q\rangle^3$ and $\langle\frac{\alpha_sGG}{\pi}\rangle\langle\bar{q}q\rangle^2\langle\bar{q}g_s\sigma Gq\rangle$ due to their  small values.
All in all, we carry out the
operator product expansion to the vacuum condensates  up to dimension-15, and  take into account the vacuum condensates $\langle\bar{q}q\rangle$, $\langle\frac{\alpha_sGG}{\pi}\rangle$, $\langle\bar{q}g_s\sigma Gq\rangle$, $\langle\bar{q}q\rangle^2$,  $\langle\bar{q}q\rangle\langle\frac{\alpha_sGG}{\pi}\rangle$,
  $\langle\bar{q} q\rangle\langle\bar{q}g_s\sigma Gq\rangle$, $\langle\bar{q} q\rangle^3$, $\langle\bar{q}q\rangle^2\langle\frac{\alpha_sGG}{\pi}\rangle$,
   $\langle\bar{q}g_s\sigma Gq\rangle^2$,  $\langle\bar{q} q\rangle^2\langle\bar{q}g_s\sigma Gq\rangle$, $\langle\bar{q}q\rangle^3\langle\frac{\alpha_sGG}{\pi}\rangle$, $\langle\bar{q} q\rangle\langle\bar{q}g_s\sigma Gq\rangle^2$,
   $\langle\bar{q}g_s\sigma Gq\rangle^3$.

Then we obtain the analytical expression of the QCD spectral density through dispersion relation,   and match the hadron side with the QCD side of the correlation function $\Pi(p^2)$  below the continuum threshold  $s_0$ and perform the Borel transformation  in regard to $P^2=-p^2$ to obtain  the   QCD sum rules:
\begin{eqnarray}\label{QCDSR-A}
\lambda^{2}_{H}\exp\left( -\frac{M_H^2}{T^2}\right)&=& \int_{9m_c^2}^{s_0}ds \,\rho_{QCD}(s)\,\exp\left( -\frac{s}{T^2}\right)\, .
\end{eqnarray}
We neglect the lengthy expression of the QCD spectral density $\rho_{QCD}(s)$ for simplicity.

We derive    Eq.\eqref{QCDSR-A} in regard to  $\tau=\frac{1}{T^2}$, then eliminate the
 pole residue $\lambda_{H}$  and obtain the QCD sum rules for
 the mass  of the triply-charmed hexaquark   state,
 \begin{eqnarray}\label{QCDSR-A-Deri}
 M^2_{H} &=& \frac{-\frac{d}{d\tau}\int_{9m_c^2}^{s_0}ds \,\rho_{QCD}(s)\,\exp\left( -s\tau\right)}{\int_{9m_c^2}^{s_0}ds \,\rho_{QCD}(s)\,\exp\left( -s\tau\right)}\, .
 \end{eqnarray}

\section{Numerical results and discussions}

We choose  the standard values of the vacuum condensates $\langle
\bar{q}q \rangle=-(0.24\pm 0.01\, \rm{GeV})^3$,   $\langle
\bar{q}g_s\sigma G q \rangle=m_0^2\langle \bar{q}q \rangle$,
$m_0^2=(0.8 \pm 0.1)\,\rm{GeV}^2$,  $\langle \frac{\alpha_s
GG}{\pi}\rangle=(0.33\,\rm{GeV})^4 $    at the energy scale  $\mu=1\, \rm{GeV}$
\cite{Reinders85,SVZ79,Colangelo-Review}, and choose the $\overline{MS}$ mass  $m_{c}(m_c)=(1.275\pm0.025)\,\rm{GeV}$
 from the Particle Data Group \cite{PDG}, and set $m_u=m_d=0$.
 We take into account
the energy-scale dependence of  the input parameters,
\begin{eqnarray}
\langle\bar{q}q \rangle(\mu)&=&\langle\bar{q}q \rangle({\rm 1GeV})\left[\frac{\alpha_{s}({\rm 1GeV})}{\alpha_{s}(\mu)}\right]^{\frac{12}{25}}\, ,\nonumber\\
\langle\bar{q}g_s \sigma Gq \rangle(\mu)&=&\langle\bar{q}g_s \sigma Gq \rangle({\rm 1GeV})\left[\frac{\alpha_{s}({\rm 1GeV})}{\alpha_{s}(\mu)}\right]^{\frac{2}{25}}\, , \nonumber\\
m_c(\mu)&=&m_c(m_c)\left[\frac{\alpha_{s}(\mu)}{\alpha_{s}(m_c)}\right]^{\frac{12}{25}} \, ,\nonumber\\
\alpha_s(\mu)&=&\frac{1}{b_0t}\left[1-\frac{b_1}{b_0^2}\frac{\log t}{t} +\frac{b_1^2(\log^2{t}-\log{t}-1)+b_0b_2}{b_0^4t^2}\right]\, ,
\end{eqnarray}
  where $t=\log \frac{\mu^2}{\Lambda^2}$, $b_0=\frac{33-2n_f}{12\pi}$, $b_1=\frac{153-19n_f}{24\pi^2}$, $b_2=\frac{2857-\frac{5033}{9}n_f+\frac{325}{27}n_f^2}{128\pi^3}$,  $\Lambda=210\,\rm{MeV}$, $292\,\rm{MeV}$  and  $332\,\rm{MeV}$ for the flavors  $n_f=5$, $4$ and $3$, respectively  \cite{PDG,Narison-mix}, and evolve all the input parameters to the pertinent  energy scale $\mu$  to extract the  mass of the triply-charmed hexaquark state with the flavor $n_f=4$.

The continuum threshold parameters are not  entirely free parameters, we often consult the experimental data to choose them. Now let us borrow some ideas from the exotic $X$, $Y$ and $Z$ states.
We usually  assign the  $Z_c^\pm(4430)$ to be the first radial excited state  of the $Z^\pm_c(3900)$ according to the
analogous decays,
\begin{eqnarray}
Z_c^\pm(3900)&\to&J/\psi\pi^\pm\, , \nonumber \\
Z_c^\pm(4430)&\to&\psi^\prime\pi^\pm\, ,
\end{eqnarray}
and the  analogous mass gaps  $M_{Z_c(4430)}-M_{Z_c(3900)}=591\,\rm{MeV}$ and $M_{\psi^\prime}-M_{J/\psi}=589\,\rm{MeV}$ from the Particle Data Group \cite{PDG,Maiani-Z4430-1405,Nielsen-1401,WangZG-Z4430-CTP}. On the other hand,  we can assign  the  $Z_c(4600)$ to be  the  vector tetraquark state with $J^{PC}=1^{--}$ \cite{Wang-Z4600-V}, or the first radial excited state of the  axialvector tetraquark state candidate $Z_c(4020)$ with $J^{PC}=1^{+-}$  \cite{ChenHX-Z4600-A,WangZG-axial-Z4600},  the energy gap between the ground state $Z_c(4020)$ and the first radial excited state $Z_c(4600)$ is about $M_{Z_c(4600)}-M_{Z_c(4020)}=576\,\rm{MeV}$ from the Particle Data Group \cite{PDG}.
In calculations, we choose the continuum threshold parameter as $\sqrt{s_0}=M_H+0.59\,\rm{GeV}\pm0.10\,\rm{GeV}$, and get a constraint to obey.

There are two basic criteria which have to be satisfied in the QCD sum rules, the one is pole dominance at the hadron  side, the other is
convergence of the operator product expansion at the QCD side.
Firstly, let us  define the pole contribution $\rm{PC}$,
\begin{eqnarray}
{\rm PC}&=& \frac{ \int_{9m_c^2}^{s_0} ds\,\rho_{QCD}(s)\,\exp\left(-\frac{s}{T^2}\right)}{\int_{9m_c^2}^{\infty} ds \,\rho_{QCD}(s)\,\exp\left(-\frac{s}{T^2}\right)}\, ,
\end{eqnarray}
and  define the contributions of the vacuum condensates of dimension  $n$,
  \begin{eqnarray}
D(n)&=& \frac{  \int_{9m_c^2}^{s_0} ds\,\rho_{QCD;n}(s)\,\exp\left(-\frac{s}{T^2}\right)}{\int_{9m_c^2}^{s_0} ds \,\rho_{QCD}(s)\,\exp\left(-\frac{s}{T^2}\right)}\, ,
\end{eqnarray}
where the $\rho_{QCD;n}(s)$ is the QCD spectral density containing the vacuum condensates of dimension $n$.

For the six-quark states, the largest power of the QCD spectral densities $\rho_{QCD}(s)\propto s^7$,
while for the four-quark states, the largest power of the QCD spectral densities $\rho_{QCD}(s)\propto s^4$,
the continuum contributions cannot be suppressed efficiently if the Borel parameters are not small enough. However, small Borel parameters lead to bad convergent behavior of the operator product expansion.   Furthermore, for the six-quark states,
the pole dominance criterion  is more difficult to satisfy compared to the cases for the four-quark states. We have to take some methods to enhance the pole contributions.

In this  article, we study the diquark-diquark-diquark type   hexaquark  states, which have  three charmed diquarks.
Such triply-charmed six-quark systems  are characterized by the effective charmed quark mass  or constituent quark mass ${\mathbb{M}}_c$
and the virtuality  $V=\sqrt{M^2_{H}-(3{\mathbb{M}}_c)^2}$, while the hidden-charm (or doubly-charmed) four-quark systems are characterized by the effective   mass   ${\mathbb{M}}_c$
and the virtuality  $V=\sqrt{M^2_{X/Y/Z}-(2{\mathbb{M}}_c)^2}$ \cite{Wang-tetra-formula,ZGWang-CTP}. We set the energy  scales of the QCD spectral densities to be $\mu=V$, it is a straight forward extension of  the energy scale formula $\mu=\sqrt{M^2_{X/Y/Z}-(2{\mathbb{M}}_c)^2}$ suggested in the QCD sum rules for the hidden-charm tetraquark  states to the triply-charmed hexaquark states \cite{Wang-tetra-formula,ZGWang-CTP}.
 In this article, we choose the updated value ${\mathbb{M}}_c=1.82\,\rm{GeV}$  \cite{Wang-EPJC-1601}, and take
 the energy scale formula,
\begin{eqnarray}\label{formula}
\mu&=&\sqrt{M^2_{H}-(3{\mathbb{M}}_c)^2}\, ,
\end{eqnarray}
 as a powerful constraint to satisfy. In previous works, we observed that the energy scale formula can enhance the pole contribution remarkably for the tetraquark (molecular) states and pentaquark (molecular) states \cite{Wang-tetra-formula,ZGWang-CTP,Wang-QQ-tetraquark,QCDSR-WangZG-5-quark-mole,QCDSR-5-quark-penta}.

Now let us  optimize  the continuum threshold parameter  $s_0$ and choose the best Borel parameter $T^2$  via trial  and error, and finally we
 obtain the Borel window, the continuum threshold parameter,  the best energy scale of the QCD spectral density, and the pole contribution, which are  shown in Table \ref{BCEPMR}.  From the Table, we can see that the pole contribution is as large as $(40-60)\%$, it is large enough to extract the hexaquark  mass reliably.

\begin{figure}
\centering
\includegraphics[totalheight=7cm,width=9cm]{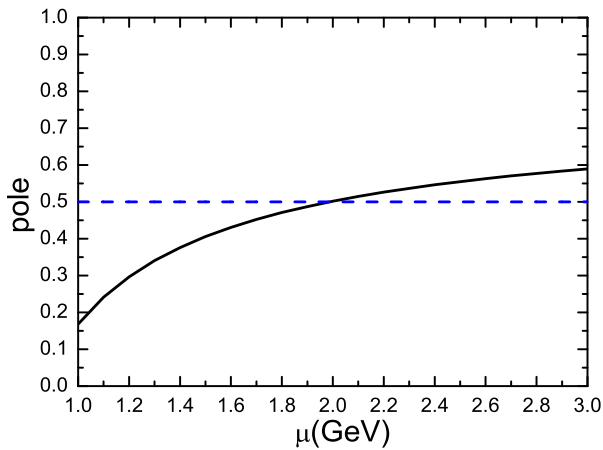}
  \caption{ The pole contribution with variation of the energy scale $\mu$ of the QCD spectral density.  }\label{Pole-mu}
\end{figure}

 We use the energy scale formula shown in Eq.\eqref{formula} to enhance the pole contribution significantly so as to satisfy the pole dominance criterion. In Fig.\ref{Pole-mu}, we plot the pole contribution with variation of the energy scale $\mu$ of the QCD spectral density for the Borel parameter $T^2=4.0\,\rm{GeV}^2$ and continuum threshold parameter $\sqrt{s_0}=6.4\,\rm{GeV}$. From the figure, we can see that
the pole contribution increases monotonously and quickly with the increase of the energy scale $\mu$ at the region $\mu<2\,\rm{GeV}$, then the pole contribution increases monotonously and slowly with the increase of the energy scale $\mu$. It is very important and necessary  to choose the pertinent energy scale $\mu$.

\begin{figure}
\centering
\includegraphics[totalheight=8cm,width=10cm]{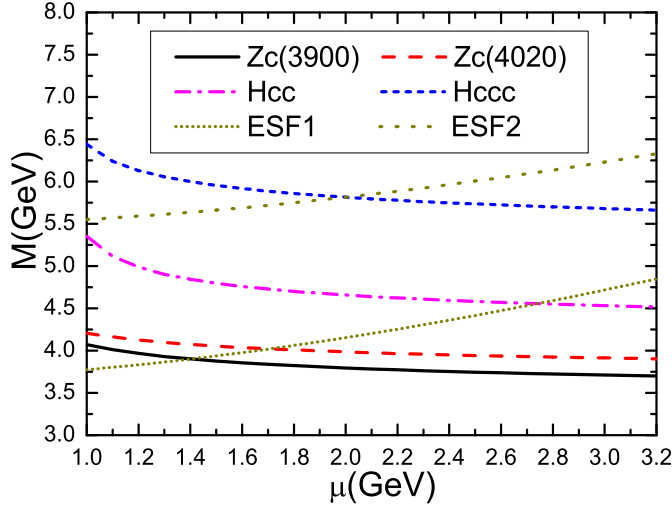}
  \caption{ The predicted masses with variations of the energy scales  $\mu$ of the QCD spectral densities, where the ESF1 and ESF2 represent the formulas
  $M=\sqrt{\mu^2+4\times (1.82\,\rm{GeV})^2}$ and $\sqrt{\mu^2+9\times (1.82\,\rm{GeV})^2}$, respectively.  }\label{mass-mu-form}
\end{figure}

We can rewrite the energy scale formulas as
\begin{eqnarray}
M_{X/Y/Z/H}&=&\sqrt{\mu^2+4{\mathbb{M}}_c^2}\, ,\nonumber\\
M_{H}&=& \sqrt{\mu^2+9{\mathbb{M}}_c^2}\, ,
\end{eqnarray}
where the $X$, $Y$, $Z$ and $H$ denote the hidden-charm or doubly-charmed or triply-charmed tetraquark states and hexaquark states.
In Fig.\ref{mass-mu-form}, we plot the predicted masses $M_{Z_c(3900)}$, $M_{Z_c(4020)}$, $M_{H_{cc}}$ and $M_{H_{ccc}}$ with variations of the energy scales $\mu$ of the QCD spectral densities, where we have taken the central values of the input parameters, see Refs.\cite{WangZG-axial-Z4600,WZG-Hidden-Charm} and Table \ref{BCEPMR}, and use the subscripts $cc$ and $ccc$ to stand for the doubly-charmed and triply-charmed hexaquark states, respectively. In Ref.\cite{WZG-hexaQ}, we construct the interpolating current $\hat{J}(x)$,
 \begin{eqnarray}
\hat{J}(x)&=&\varepsilon^{abc}\varepsilon^{aij}\varepsilon^{bkl}\varepsilon^{cmn}\, u^{T}_i(x)C\gamma_5d_j(x) \,u^{T}_k(x)C\gamma_5 c_l(x)\, d^{T}_m(x)C\gamma_5 c_n(x) \, ,
\end{eqnarray}
 to study the scalar-diquark-scalar-diquark-scalar-diquark  type doubly-charmed hexaquark state $uuddcc$ or $H_{cc}$ with the QCD sum rules. After the article was published, we checked the calculations again and observed that the QCD spectral densities $\rho_{i}(s)$ with $i=3$, $5$,  $9$, $11$ and $13$ should change a minus sign, $\rho_{i}(s)\to -\rho_i(s)$. Now we recalculate the mass of the  doubly-charmed hexaquark state $uuddcc$ or $H_{cc}$ with all the updated parameters in a consistent  way, the relevant parameters and the numerical results  are also presented in Table \ref{BCEPMR}.

 From Fig.\ref{mass-mu-form}, we can see that the predicted masses of the tetraquark states and hexaquark states decrease monotonically with the increase of the energy scales  $\mu$, the  line $M=\sqrt{\mu^2+4\times (1.82\,\rm{GeV})^2}$ intersects with the lines of the masses of the $Z_c(3900)$, $Z_c(4020)$ and $H_{cc}$ tetraquark or hexaquark states at the
energy scales about $ \mu= 1.4\,\rm{GeV}$, $1.7\,\rm{ GeV}$ and $2.7\,\rm{GeV}$, respectively, which happen to reproduce the experimental values of the masses of the $Z_c(3900)$ and $Z_c(4020)$,  respectively. The energy scale formula serves a milestone to choose the pertinent energy scales of the QCD spectral densities. Accordingly, the line $M=\sqrt{\mu^2+9\times (1.82\,\rm{GeV})^2}$ intersects with the line of the mass of the triply-charmed hexaquark state  $H_{ccc}$ at the energy scale $\mu=2.0\,\rm{GeV}$, which is expected to
be the pertinent energy scale of the QCD spectral density and leads to the ideal mass.
The mass gap $M_{H_{ccc}}-M_{H_{cc}}=1.25\,\rm{GeV}$ happens to be the $\overline{MS}$ mass of the $c$-quark, $m_c(m_c)$. We can draw the conclusion tentatively that
the energy scale  formula can be applied to study the hexaquark states in a consistent way.

\begin{figure}
\centering
\includegraphics[totalheight=7cm,width=9cm]{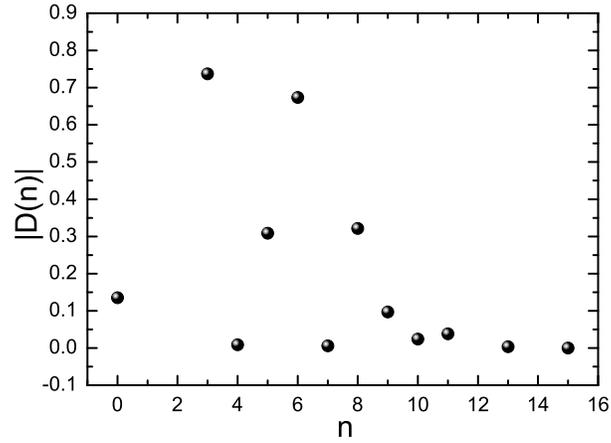}
  \caption{ The absolute values of the contributions of the vacuum condensates of dimension $n$ for  central values of the input parameters.  }\label{OPE}
\end{figure}

In Fig.\ref{OPE}, we plot the absolute values of the $D(n)$ for the central values of the input parameters shown in Table \ref{BCEPMR}.
From the figure, we can see that the contributions of the vacuum condensates with the dimensions $n\leq 8$ vibrate, the contribution of the perturbative term or $D(0)$ is  small, the contributions $D(3)$ and $D(6)$ are very large, the contributions $D(4)$ and $D(7)$ are tiny,  however, such vibrations cannot destroy the convergence of the operator product expansion. The vacuum condensate $\langle\bar{q}q\rangle\langle\bar{q}g_s\sigma Gq\rangle$ with the dimension $8$ serves   as a milestone,   the absolute values of the  contributions $|D(n)|$ with $n\geq 8$ decrease  monotonically  and quickly with the increase of the dimensions $n$, the value $|D(15)|\approx 0$, the operator product expansion is convergent.

\begin{figure}
\centering
\includegraphics[totalheight=7cm,width=9cm]{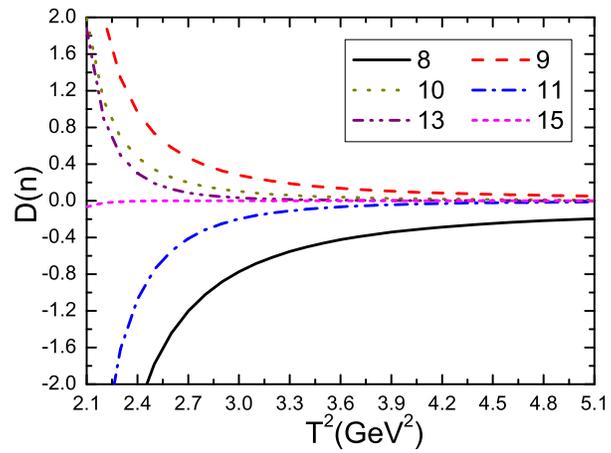}
  \caption{ The  contributions of the higher dimensional vacuum condensates with variation of the Borel parameter $T^2$.  }\label{OPE-high}
\end{figure}

In Fig.\ref{OPE-high}, we plot the contributions of the vacuum condensates  $D(n)$ with $n\geq 8$ for the central values of the input parameters shown in Table \ref{BCEPMR}.  From the figure, we can see that the contributions of the higher dimensional vacuum condensates decrease monotonously and quickly  with the
 increase of the Borel parameter $T^2$ at the region $T^2\leq 3.3\,\rm{GeV}^2$, then  they decrease monotonously and slowly with the increase of the Borel parameter $T^2$.
It is reasonable to choose the value $T^2>3.3\,\rm{GeV}^2$.  The higher dimensional vacuum condensates play a minor important role in the Borel windows, but they play an important role in determining  the Borel windows.

\begin{figure}
\centering
\includegraphics[totalheight=8cm,width=10cm]{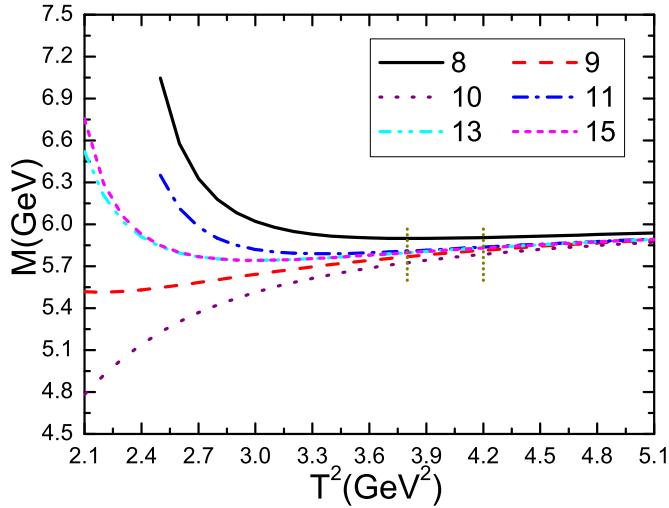}
  \caption{ The predicted masses with variation of the Borel parameter $T^2$ for the truncations of the operator product expansion up to the vacuum condensates of dimensions $n=8$, $9$, $10$, $11$, $13$ and $15$, the region between the two vertical lines is the Borel window.  }\label{mass-ccc-dim}
\end{figure}

In Fig.\ref{mass-ccc-dim}, we plot the predicted triply-charmed hexaquark masses with variation of the Borel parameter $T^2$ for the truncations of the operator product expansion up to the vacuum condensates of dimensions $n=8$, $9$, $10$, $11$, $13$ and $15$ respectively with the central values of the other parameters shown in Table \ref{BCEPMR}.  From the figure,  we can see that the predicted masses change greatly  with the truncations  $n\leq 13$ at the region $T^2<3.8\,\rm{GeV}^2$, which is consistent with the behavior of the operator product expansion,  the contributions of the higher dimensional vacuum condensates are large and change greatly at the region $T^2\leq 3.3\,\rm{GeV}^2$. In this region, we cannot obtain  flat platforms.  If we take the truncations $n\geq 9$, the predicted masses change slightly in the Borel window $T^2=(3.8-4.2)\,\rm{GeV}^2$. However, the higher dimensional vacuum condensates play an important role in determining  the Borel windows, without taking into account the vacuum condensates up to dimension $15$, we cannot obtain the Borel window $T^2=(3.8-4.2)\,\rm{GeV}^2$, although they  play a minor important role in the Borel window.

\begin{figure}
\centering
\includegraphics[totalheight=6cm,width=7cm]{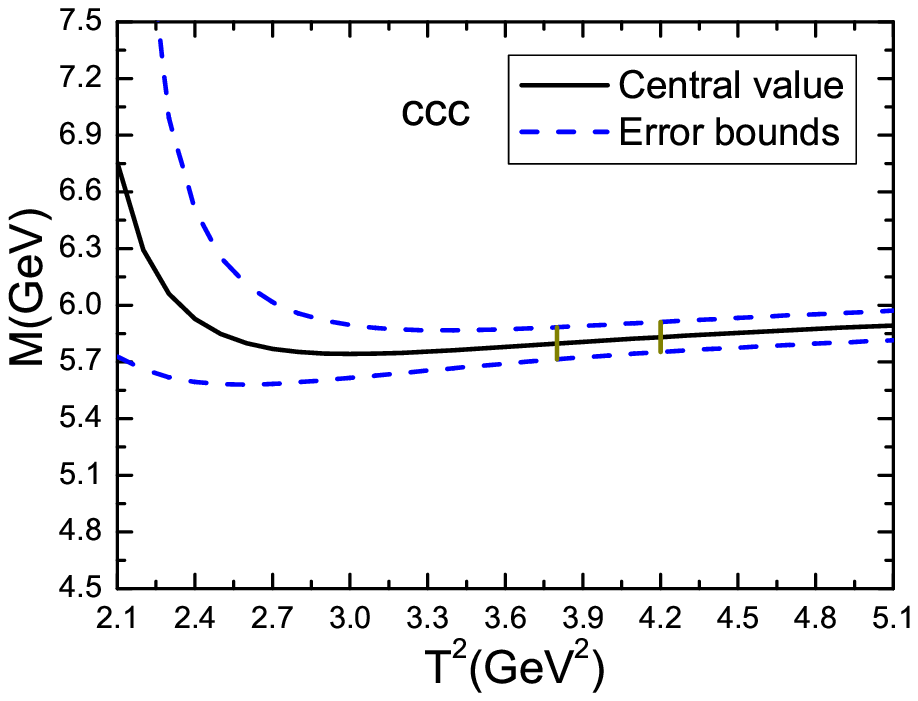}
\includegraphics[totalheight=6cm,width=7cm]{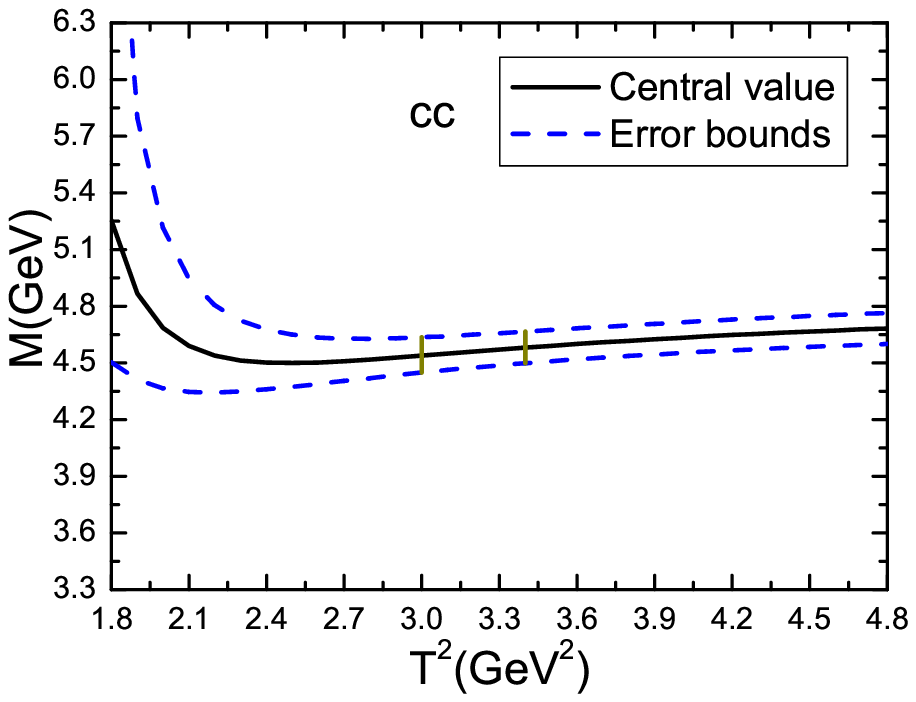}
  \caption{ The masses of the triply-charmed and doubly-charmed hexaquark states with variation of the Borel parameter $T^2$, the regions  between the two vertical lines are the Borel windows.  }\label{mass-ccc}
\end{figure}

\begin{figure}
\centering
\includegraphics[totalheight=6cm,width=7cm]{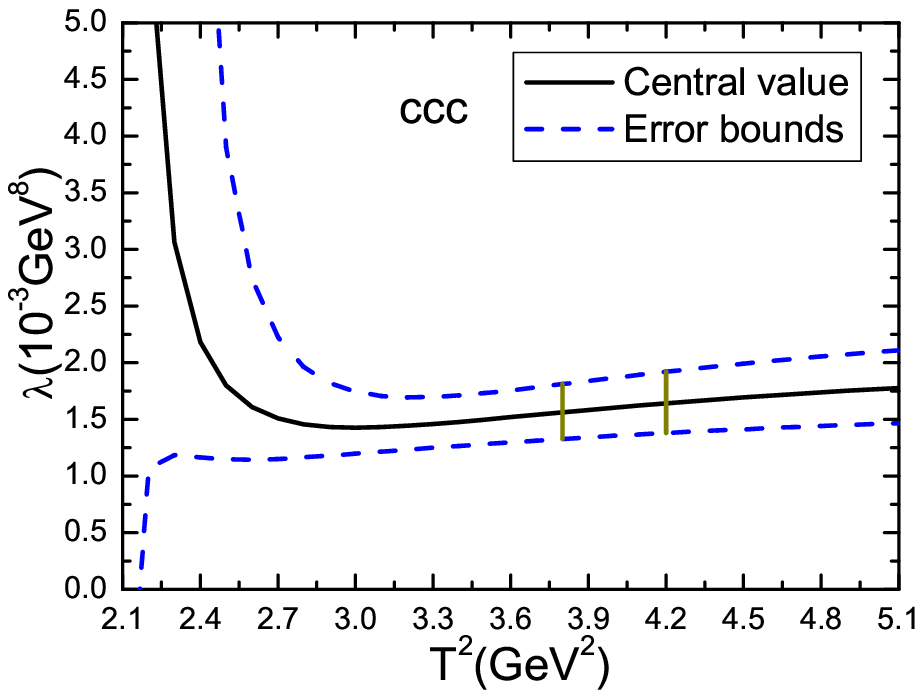}
\includegraphics[totalheight=6cm,width=7cm]{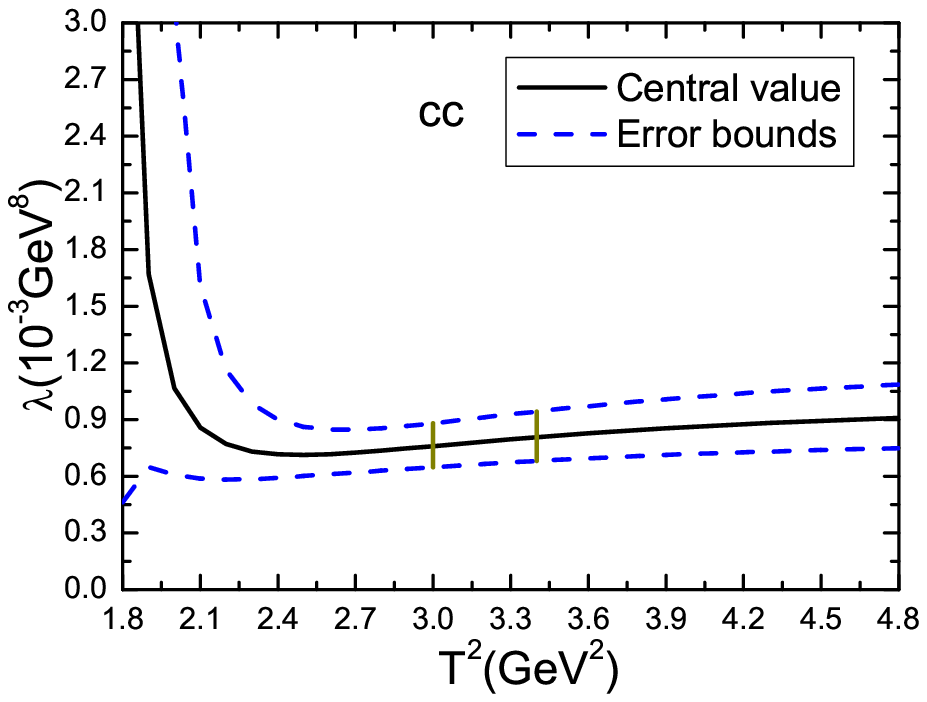}
  \caption{  The pole residues  of the triply-charmed and doubly-charmed hexaquark states with variation of the Borel parameter $T^2$, the  regions  between the two vertical lines are the Borel windows.  }\label{residue-ccc}
\end{figure}

Now we take  into account all uncertainties of the input parameters,
and obtain the values of the mass and pole residue of
 the   triply-charmed hexaquark state, which are  shown explicitly in Table \ref{BCEPMR} and Figs.\ref{mass-ccc}-\ref{residue-ccc}, where we also present
 the results for the scalar doubly-charmed hexaquark state.

 From Figs.\ref{mass-ccc}-\ref{residue-ccc}, we can see that there appear flat platforms  in the Borel windows both for the masses and pole residues, it is reliable to extract the hexaquark masses. From Table \ref{BCEPMR}, we can see that the central values of the hexaquark masses  satisfy the energy scale formula  $\mu=\sqrt{M^2_{H}-(3{\mathbb{M}}_c)^2}$ and $\sqrt{M^2_{H}-(2{\mathbb{M}}_c)^2}$, respectively.  Also from Table \ref{BCEPMR}, we can  obtain reasonably relation between the doubly-charmed and triply-charmed hexaquark states, $M_{H_{ccc}}-M_{H_{cc}}=m_c(m_c)$, which valuates  the present calculations.

\begin{table}
\begin{center}
\begin{tabular}{|c|c|c|c|c|c|c|c|}\hline\hline
$J^P$          &$T^2(\rm{GeV}^2)$   &$\sqrt{s_0}(\rm{GeV})$   &$\mu(\rm{GeV})$  &pole          &$M(\rm{GeV})$  &$\lambda(10^{-3}\rm{GeV}^8)$ \\ \hline

$1^+(cccuud)$  &$3.8-4.2$           &$6.40\pm0.10$            &$2.0$            &$(40-60)\%$   &$5.81\pm0.10$  &$1.60\pm0.30$  \\ \hline

$0^+(ccuudd)$  &$3.0-3.4$           &$5.15\pm0.10$            &$2.7$            &$(40-63)\%$   &$4.56\pm0.11$  &$0.78\pm0.15$  \\ \hline\hline
\end{tabular}
\end{center}
\caption{ The Borel parameters, continuum threshold parameters,   energy scales, pole contributions,   masses and pole residues for the
 triply-charmed  and doubly-charmed hexaquark   states. }\label{BCEPMR}
\end{table}

In Ref.\cite{WZG-ccc-dibaryon}, we  construct the color-singlet-color-singlet type currents to study the scalar and axialvector $\Xi_{cc}\Sigma_c$ dibaryon states with QCD sum rules, and obtain the masses $M_{\Xi_{cc}\Sigma_c(0^+)}=6.05\pm0.13\,\rm{GeV}$ and $M_{\Xi_{cc}\Sigma_c(1^+)}=6.03\pm0.13\,\rm{GeV}$, which lie above mass of the diquark-diquark-diquark type triply-charmed hexaquark state, $M_{H_{ccc}(1^+)}=5.81\pm0.10\,\rm{GeV}$.
In Ref.\cite{WZG-ccc-dibaryon}, we construct the current $\tilde{J}_\mu(x)$ to interpolate the axialvector $\Xi_{cc}\Sigma_c$ dibaryon state,
\begin{eqnarray}\label{Dibaryon-current}
\tilde{J}_\mu(x)&=& J^T_{c}(x) C\gamma_\mu J_{cc}(x)\, ,
\end{eqnarray}
where
\begin{eqnarray}
 J_{c}(x) &=&   \varepsilon^{ijk}q^T_i(x)C\gamma_\mu q_j(x) \gamma^\mu\gamma_5c_k(x) \, , \nonumber\\
 J_{cc}(x) &=&   \varepsilon^{ijk}c^T_i(x)C\gamma_\mu c_j(x) \gamma^\mu\gamma_5q_k (x)\, .
  \end{eqnarray}
Now let us perform the Fierz-rearrangements for the currents $\tilde{J}_\mu(x)$ and $J_\mu(x)$ both in the Dirac spinor space and color space to obtain the results,
\begin{eqnarray}\label{Fierz-dibaryon}
 J_{c} &=&-\varepsilon^{ijk}q^T_i C c_j \gamma_5q_k+\varepsilon^{ijk}q^T_iC\gamma_5 c_j q_k-\frac{1}{2}\varepsilon^{ijk}q^T_iC\gamma_\alpha c_j \gamma^{\alpha}\gamma_5q_k
 -\frac{1}{2}\varepsilon^{ijk}q^T_iC\gamma_\alpha\gamma_5 c_j \gamma^{\alpha}q_k\, ,  \nonumber\\
 J_{cc}  &=&-\varepsilon^{ijk}c^T_i C q_j \gamma_5c_k+\varepsilon^{ijk}c^T_iC\gamma_5 q_j c_k-\frac{1}{2}\varepsilon^{ijk}c^T_iC\gamma_\alpha q_j \gamma^{\alpha}\gamma_5c_k
 -\frac{1}{2}\varepsilon^{ijk}c^T_iC\gamma_\alpha\gamma_5 q_j \gamma^{\alpha}c_k\, , \nonumber\\
  \end{eqnarray}
\begin{eqnarray}\label{Fierz-Hexquark}
 J_\mu &=&   \left[\varepsilon^{ijm}u^T_iC\gamma_5c_ju^T_m\right]C\gamma_{\mu} \left[\varepsilon^{kln} d^T_kC\gamma_5c_l  c_n \right] +\left[\varepsilon^{ijn}u^T_iC\gamma_5c_jc^T_n \right]C\gamma_{\mu} \left[ \varepsilon^{klm}d^T_kC\gamma_5c_l u_m \right]\, . \nonumber\\
  \end{eqnarray}
From Eq.\eqref{Dibaryon-current} and Eqs.\eqref{Fierz-dibaryon}-\eqref{Fierz-Hexquark}, we can observe that there are scalar, pseudoscalar,  axialvector and vector charmed diquark operators in the current  $\tilde{J}_\mu(x)$, while there are only scalar charmed diquark operators  in the current $J_\mu(x)$.
The favored quark-quark configurations are the scalar and axialvector diquark states from the QCD sum rules,  the axialvector diquark states have slightly larger masses than the corresponding scalar diquark states \cite{WangDiquark,WangLDiquark}. It is natural  that the diquark-diquark-diquark type triply-charmed hexaquark state has small mass than the color-singlet-color-singlet type  $\Xi_{cc}\Sigma_c$ dibaryon state.
The decays of the triply-charmed hexaquark state $H_{ccc}$ to the final states  $\Xi_{cc}\Sigma_c$, $\Xi_{cc}\Lambda_c$ and $\Omega_{ccc}p$ are kinematically  forbidden \cite{PDG,WangZG-ccc-CTP}. As the $\Omega_{ccc}$ has not been observed yet, we choose the mass from the theoretical calculations \cite{WangZG-ccc-CTP}. The triply-charmed hexaquark state $H_{ccc}$ decays   weakly, the width is expected to be small. If the spin-breaking effects are small,  the ground state hexaquark states with the spin-parity $J^P=0^+$, $1^+$ and $2^+$ have almost degenerated   masses, and analogous narrow widths. We  can search for the triply-charmed and doubly-charmed  hexaquark states  at the LHCb, Belle II,  CEPC, FCC, ILC  in the future.

\section{Conclusion}
In this article,  we construct the diquark-diquark-diquark type current to interpolate the  triply-charmed axialvector hexaquark state, and study its mass and pole residue with the QCD sum rules in details by carrying out the operator product expansion up to the vacuum condensates of dimension 15. In calculations, we choose the best energy scale of the QCD spectral density  with the  energy scale formula $\mu=\sqrt{M^2_{H}-(3{\mathbb{M}}_c)^2}$, which can enhance the pole contribution remarkably to satisfy the pole dominance criterion at the hadron side and improve the convergent behavior of the operator product expansion by suppressing  the contributions of the higher dimensional vacuum condensates  at the QCD side.
 Finally, we obtain the mass and pole residue of the triply-charmed axialvector  hexaquark state, the predicted mass lies below the two-baryon thresholds, which indicates that the triply-charmed hexaquark state $H_{ccc}$ decays weakly, the width is small. The results  should be taken with caution, as we could  obtain more robust  predictions  by choosing the most general currents. If the spin-breaking effects are small,  the ground state hexaquark states with the spin-parity  $J^P=0^+$, $1^+$ and $2^+$ have almost degenerated masses, and analogous narrow widths.  Furthermore, we re-analyze   the mass of the diquark-diquark-diquark type scalar doubly-charmed hexaquark state, and obtain reasonable  relation between the doubly-charmed and triply-charmed hexaquark states.
We  can search for the triply-charmed and doubly-charmed hexaquark  states  at the LHCb, Belle II,  CEPC, FCC, ILC  in the future.

\section*{Acknowledgements}
This  work is supported by National Natural Science Foundation, Grant Number  11775079.


\begin{thebibliography}{99}

\bibitem{X3872-2003} S. K. Choi   et al, Phys. Rev. Lett. {\bf 91} (2003) 262001.

\bibitem{HXChen-review-1601} H. X. Chen, W. Chen, X. Liu  and S. L. Zhu,  Phys. Rept. {\bf 639} (2016) 1.

\bibitem{RFLebed-review-1610} R. F. Lebed, R. E. Mitchell and E. S. Swanson, Prog. Part. Nucl. Phys. {\bf 93} (2017) 143.


\bibitem{AEsposito-review-1611} A. Esposito, A. Pilloni and A. D. Polosa,  Phys. Rept. {\bf 668} (2017) 1.


\bibitem{FKGuo-review-1705}  F. K. Guo, C. Hanhart, U. G. Meissner, Q. Wang, Q. Zhao and B. S. Zou, Rev. Mod. Phys. {\bf 90} (2018) 015004.


\bibitem{AAli-review-1706} A. Ali, J. S. Lange and S. Stone, Prog. Part. Nucl. Phys. {\bf 97} (2017) 123.

\bibitem{SLOlsen-review-1708} S. L. Olsen, T. Skwarnicki and D. Zieminska, Rev. Mod. Phys. {\bf 90} (2018)  015003.

\bibitem{MNielsen-review-1812} R. M. Albuquerque, J. M. Dias, K. P. Khemchandani, A. M. Torres, F. S. Navarra, M. Nielsen and C. M. Zanetti,
J. Phys. {\bf G46} (2019) 093002.


\bibitem{YRLiu-review-1903} Y. R. Liu, H. X. Chen, W. Chen, X. Liu and S. L. Zhu, Prog. Part. Nucl. Phys. {\bf 107} (2019) 237.


\bibitem{CPShen-review-1907} N. Brambilla, S. Eidelman, C. Hanhart, A. Nefediev, C. P. Shen, C. E. Thomas, A. Vairo and C. Z. Yuan, arXiv:1907.07583.



\bibitem{QCDSR-4-quark-mass} R. D. Matheus, S. Narison, M. Nielsen and J. M. Richard, Phys. Rev. { \bf D75} (2007) 014005;
S. H. Lee, A. Mihara, F. S. Navarra and M. Nielsen, Phys. Lett. {\bf B661} (2008) 28;
Z. G. Wang, Eur. Phys. J. {\bf C63} (2009) 115;
W. Chen and S. L. Zhu, Phys. Rev. {\bf D83} (2011) 034010;
J. R. Zhang, M. Zhong and M. Q. Huang, Phys. Lett. {\bf B704} (2011) 312;
J. R. Zhang,  Phys. Rev. {\bf D87} (2013)  116004;
C. F. Qiao and L. Tang, Eur. Phys. J. {\bf C74} (2014) 2810;
S. S. Agaev, K. Azizi and H. Sundu, Eur. Phys. J. {\bf C77} (2017)  321;
Z. G. Wang,  Eur. Phys. J. {\bf C79} (2019)  29;
Z. G. Wang, Commun. Theor. Phys. {\bf 71} (2019)  1319.




\bibitem{WangHuangtao-PRD} Z. G. Wang and T. Huang, Phys. Rev. {\bf D89} (2014)  054019.

\bibitem{Wang-tetra-formula}  Z. G. Wang, Eur. Phys. J. {\bf C74} (2014)  2874.


\bibitem{ZGWang-CTP} Z. G. Wang, Commun. Theor. Phys. {\bf 63} (2015) 466;
Z. G. Wang and Y. F. Tian, Int. J. Mod. Phys. {\bf A30} (2015) 1550004.


\bibitem{QCDSR-4-quark-width} F. S. Navarra and M. Nielsen, Phys. Lett. {\bf B639} (2006) 272;
J. M. Dias, F. S. Navarra, M. Nielsen, C. M. Zanetti, Phys. Rev. {\bf D88} (2013)  016004;
W. Chen, T. G. Steele, H. X. Chen and S. L. Zhu, Eur. Phys. J. {\bf C75} (2015)  358;
Z. G. Wang and T. Huang, Nucl. Phys. {\bf A930} (2014) 63;
S. S. Agaev, K. Azizi and H. Sundu, Phys. Rev. {\bf D93} (2016)  074002;
Z. G. Wang and J. X. Zhang, Eur. Phys. J. {\bf C78} (2018)  14;
Z. G. Wang and Z. Y. Di, Eur. Phys. J. {\bf C79} (2019) 72.



\bibitem{WangZG-4-quark-mole} Z. G. Wang and T. Huang, Eur. Phys. J. {\bf C74} (2014)  2891;
Z. G. Wang, Eur. Phys. J. {\bf C74} (2014)  2963.


\bibitem{WangDiquark} Z. G. Wang, Eur. Phys. J. {\bf C71} (2011) 1524;
  R. T. Kleiv, T. G. Steele and A. Zhang, Phys. Rev. {\bf D87} (2013) 125018.

\bibitem{WangLDiquark}  Z. G. Wang, Commun. Theor. Phys. {\bf 59} (2013) 451.



\bibitem{LHCb-Xicc} R. Aaij  et al,  Phys. Rev. Lett. {\bf 119} (2017)  112001.


\bibitem{Nielsen-Lee} F. S. Navarra, M. Nielsen and S. H. Lee, Phys. Lett. {\bf B649} (2007) 166.

\bibitem{QQ-QCDSR-Wang2011} Z. G. Wang, Y. M. Xu and H. J. Wang, Commun. Theor. Phys. {\bf 55} (2011) 1049.

\bibitem{QQ-QCDSR-Chen} M. L. Du, W. Chen, X. L. Chen and S. L. Zhu,  Phys. Rev. {\bf D87} (2013) 014003.


\bibitem{Wang-QQ-tetraquark} Z. G. Wang,  Acta Phys. Polon. {\bf B49} (2018) 1781;
Z. G. Wang and Z. H. Yan, Eur. Phys. J. {\bf C78} (2018) 19.



\bibitem{WZG-hexaQ} Z. G. Wang, Eur. Phys. J. {\bf C77} (2017)  642.


\bibitem{WZG-ccc-dibaryon}  Z. G. Wang, arXiv:1912.07230.


\bibitem{Reinders85} L. J. Reinders, H. Rubinstein and S. Yazaki, Phys. Rept. {\bf 127} (1985) 1.


\bibitem{Pascual-1984} P. Pascual and R. Tarrach, ``QCD: Renormalization for the practitioner", Springer Berlin Heidelberg (1984).


\bibitem{QCDSR-WangZG-5-quark-mole}  Z. G. Wang, Int. J. Mod. Phys. {\bf A34} (2019)  1950097.


\bibitem{QCDSR-5-quark-penta} Z. G. Wang, Eur. Phys. J. {\bf C76} (2016)  70;
Z. G.  Wang and T. Huang, Eur. Phys. J. {\bf C76} (2016)  43.




\bibitem{SVZ79} M. A. Shifman, A. I. Vainshtein and V. I. Zakharov, Nucl. Phys. {\bf B147} (1979) 385; Nucl. Phys. {\bf B147} (1979) 448.


\bibitem{Colangelo-Review}  P. Colangelo and A. Khodjamirian, hep-ph/0010175.

\bibitem{PDG}  M. Tanabashi et al, Phys. Rev. {\bf  D98} (2018)  030001.


\bibitem{Narison-mix} S. Narison and R. Tarrach, Phys. Lett. {\bf 125 B} (1983) 217.


\bibitem{Maiani-Z4430-1405} L. Maiani, F. Piccinini, A. D. Polosa and V. Riquer, Phys. Rev. {\bf D89} (2014) 114010.


\bibitem{Nielsen-1401} M. Nielsen and F. S. Navarra,  Mod. Phys. Lett. {\bf  A29} (2014) 1430005.

\bibitem{WangZG-Z4430-CTP} Z. G. Wang,  Commun. Theor. Phys. {\bf 63} (2015)  325.



\bibitem{Wang-Z4600-V} Z. G. Wang, Int. J. Mod. Phys. {\bf A34} (2019)  1950110.

\bibitem{ChenHX-Z4600-A} H. X. Chen and W. Chen,  Phys. Rev. {\bf D99} (2019)  074022.

\bibitem{WangZG-axial-Z4600} Z. G. Wang, Chin. Phys. {\bf C44} (2020) 063105.


\bibitem{Wang-EPJC-1601} Z. G. Wang,  Eur. Phys. J. {\bf C76} (2016)  387.



\bibitem{WZG-Hidden-Charm} Z. G. Wang, arXiv:1908.07914.



\bibitem{WangZG-ccc-CTP} Z. G. Wang,  Commun. Theor. Phys. {\bf 58} (2012) 723;
M. S. Liu, Q. F. Lu and X. H. Zhong, arXiv:1912.11805.

\end{thebibliography}
\end{document}